# Reconstruction of low dimensional electronic states by altering the chemical arrangement at the SrTiO$_3$ surface

Hang Li[1,2] *, Walber H. Brito[3], Eduardo B. Guedes[2], Alla Chikina[2], Rasmus T. Dahm[1], Dennis V. Christensen[1], Shinhee Yun[1], Francesco M. Chiabrera[1], Nicholas C. Plumb[2], Ming Shi[2], Nini Pryds[1] and Milan Radovic[1,2] #

[1]*Department of Energy Conversion and Storage, Technical University of Denmark, 2800 Kgs. Lyngby, Denmark*

[2]*Photon Science Division, Paul Scherrer Institute, 5232 Villigen-PSI, Switzerland*

[3]*Departamento de Física, Universidade Federal de Minas Gerais, C. P. 702, 30123-970 Belo Horizonte, Minas Gerais, Brazil*

Email: * hang.li@psi.ch, # milan.radovic@psi.ch

## Abstract

Developing reliable methods for modulating the electronic structure of the two-dimensional electron gas (2DEG) in SrTiO$_3$ is crucial for utilizing its full potential and inducing novel properties. Here, we show that relatively simple surface preparation reconstructs the 2DEG of SrTiO$_3$ (STO) surface, leading to a Lifshitz-like transition. Combining experimental methods, such as angle-resolved photoemission spectroscopy (ARPES) and X-ray photoemission spectroscopy (XPS) with *ab initio* calculations, we find that the modulation of the surface band structures is primarily attributed to the reorganization of the chemical composition. In addition, ARPES experiments demonstrate that vacuum ultraviolet (VUV) light can be efficiently employed to alter the band renormalization of the 2DEG system and control the electron-phonon interaction (EPI). Our study provides a robust and straightforward route to stabilize and tune the low-dimensional electronic structure via the chemical degeneracy of the STO surface.

## Introduction

Transition metal oxide-based interfaces and surfaces, in particular those based on STO, exhibit a plethora of properties such as superconductivity [1-5], magnetism [6-9], Rashba-type spin-orbital coupling [10,11], and quantum Hall effect [12,13]. SrTiO$_3$, with a cubic perovskite structure, is a typical choice as a substrate for epitaxial growth of many oxides. In cubic STO, the octahedral crystal field splits the Ti $3d$ orbitals in the well-known $t_{2g}$ and $e_g$ subbands, with degenerate $t_{2g}$ states at the $\Gamma$ point and the $e_g$ states lying at higher energies. Structural relaxation and reconstructions at the STO surface and interface regions can lift the degeneracy and lower the dimensionality of the electronic bands [14]. The evolution of the electronic phases in STO-based systems and its relation to the properties have been intensively investigated: e.g., high mobility in γ-$Al_2O_3$/STO [15,16], anomalous Hall effect induced by Lifshitz transition [17], and quantum Hall effect [12,13], Rashba-like spin structure [18, 19]. The mentioned complex phenomena, which often coincide, demonstrate that understanding the electronic structure and the low degeneracy surface is crucial to understanding the causes of these behaviors and achieving control over them. Because of its simplicity, studies on the bare $SrTiO_3$ surface are essential to shed light on the fundamental mechanisms leading to the observed band order and predicting new ways for their manipulation.

The band modulations in bare STO can be modified by temperature change [20-23], stress [24-26], and surface termination [27-29]. Angle-resolved photoemission spectroscopy (ARPES) studies on nominally $TiO_2$-terminated STO single crystal [14, 30-32] display the typical electronic structures of STO consisting of shallow $d_{xz}/d_{yz}$ bands and deep $d_{xy}$ subbands – a fingerprint of most STO-based systems. Interestingly, a single band was observed in epitaxial-grown SrO layer on $TiO_2$-terminated STO [27]. These studies suggest that the surface termination and its chemical composition plays a crucial role in determining the properties of the underlying 2DEG.

Figure 1 shows a schematic illustration of two possible surfaces terminations of STO (001) ($TiO_2$ and SrO) and their combination (panel 1a) as well as their electronic structures (panel 1c) [14,18,26,27,30,31]. If the octahedral symmetry of the 001-oriented STO is preserved at the surface, it protects the degeneracy of the $t_{2g}$ ($d_{xy}$, $d_{xz}$ and $d_{yz}$) bands. Lowering the symmetry from octahedral to tetragonal leads to splitting of the $d_{xy}$ and the $d_{xz}/d_{yz}$ bands (Figure 1b), characterized by the energy difference, $\Delta t_{2g}$. We have used this parameter to evaluate the lowering of dimensionality and the degeneracy of Nb-doped STO wafers through surface engineering. Furthermore, it was shown that increasing temperature causes a depletion of the 2DEG of STO [23] while reducing the orbital splitting, which can be further controlled by strain [26].

By combining Ar sputtering and vacuum annealing, we establish a vigorous procedure to tune the 2DEG of STO (Fig.1c). Importantly, we report that the 2D $d_{xy}$ subband emerges at a Sr-enriched surface, leading to a pure single-orbital 2DEG system. This electronic phase is found to be air-stable, offering an advantage for applications. Meanwhile, the band splitting between $d_{xy}$ and $d_{xz}/d_{yz}$ caused by ultraviolet (VUV) irradiations demonstrates that light can be used as a knob for tuning the band splitting ($\Delta t_{2g}$) and Lifshitz-like transition in the 2DEG system.

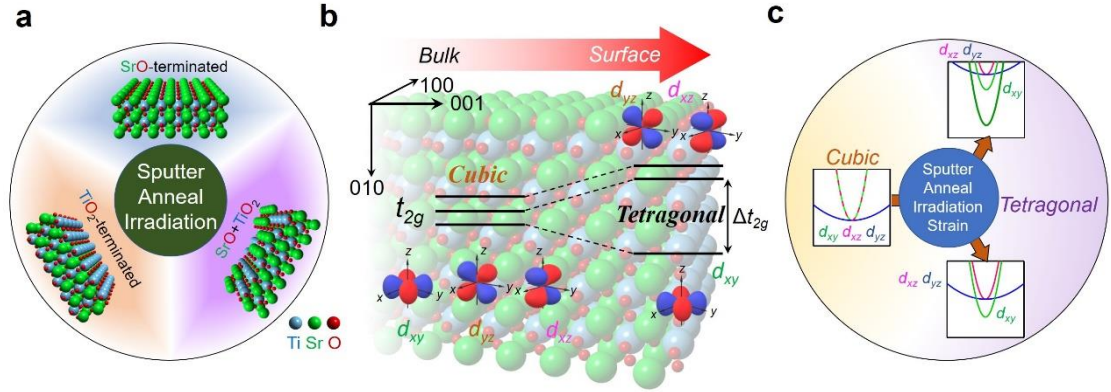

Figure 1. **a**, Schematic illustration of SrTiO$_3$ surfaces with three types of terminations, SrO-termination, TiO$_2$-termination and SrO+TiO$_2$-termination, respectively. **b**, Tetragonal crystal field splitting between $d_{xy}$ and $d_{xz}/d_{yz}$ bands, where $\Delta t_{2g}$ marks the energy difference. **c**, Schematic illustration of the electronic structures of STO surfaces observed in our study.

## Results and discussion

### Band structures of surface-engineered STO

Figure 2 displays the electronic structures of STO single crystal (001) surfaces measured by ARPES after various preparation processes. The as-received STO samples are nominally TiO$_2$-terminated with 0.5% Nb doping (SurfaceNet GmbH). The characterizations of the "as-received" sample (stage #1) were performed by ARPES and XPS and used as reference. Subsequently, the sample was treated by Ar sputtering and annealing (stage #2: 5 minutes of Ar sputtering followed by annealing in ultra-high vacuum (UHV) at 700 °C for one hour). Afterward, we annealed the sample at 800 °C for 2 hours in UHV (stage #3). Both stages (#2 and #3) were studied by XPS and ARPES. The detailed surface preparation procedure is presented in the Methods section and in figure S1.

The ARPES data in Figure 2 were obtained using circular polarized (C+) light.

With such light polarization, both in-plane $d_{xy}$ and out-of-plane $d_{xz}/d_{yz}$ orbitals at the STO surface are probed [20-22], showing the evolution of the electronic structure during irradiation. Figures 2a-c display the electronic structures of STO in stages #1, #2, and #3 after 100 minutes of irradiation along the $\Gamma$-$X$ crystal direction. Similar to previous ARPES studies on STO surfaces [14, 30-32], the electronic structures of as-received STO (#1) and high-temperature annealed STO (#3) after irradiation of $t_f$~1.6 h show degenerate $d_{xz}$ and $d_{yz}$ bands and down-shifted $d_{xy}$ subbands (Figure 2a,c). The band structure measurement of stage #2 (Figure 2b) shows that only one $d_{xy}$ subband near the $\Gamma$ point is occupied. This indicates an electronic transition from a multi-band (#1) to a single-band system (#2) and back (#3).

The band character of stages #1 to #3 is depicted by in-plane Fermi surfaces maps and $k_z$ maps (Figure 2j-l). For stages #1 and #3 (Figure 2 j,l), the in-plane Fermi surfaces (FSs) consist of one circular electron pocket and two intersecting ellipsoidal electron pockets centered at $\Gamma$, which are typical for the STO (001) surface [14, 30, 31]. The $d_{yz}$ and $d_{xz}$ bands for stages #1 and #3 exhibit quasi-3D characters, while the $d_{xy}$ band shows two-dimensional (2D) character (Figure 2 j,l). In contrast, stage #2 is characterized only by a single circular electron pocket around $\Gamma$ (Figure 2k) with 2D character – i.e., without dispersion in the $k_z$ direction (Figure 2k). The occupation of a single band was earlier reported in the LAO/STO [17] and STO systems [19, 27]. However, in the following section, we will discuss the origin and properties of the single $d_{xy}$ band with pure 2D character found in stage #2.

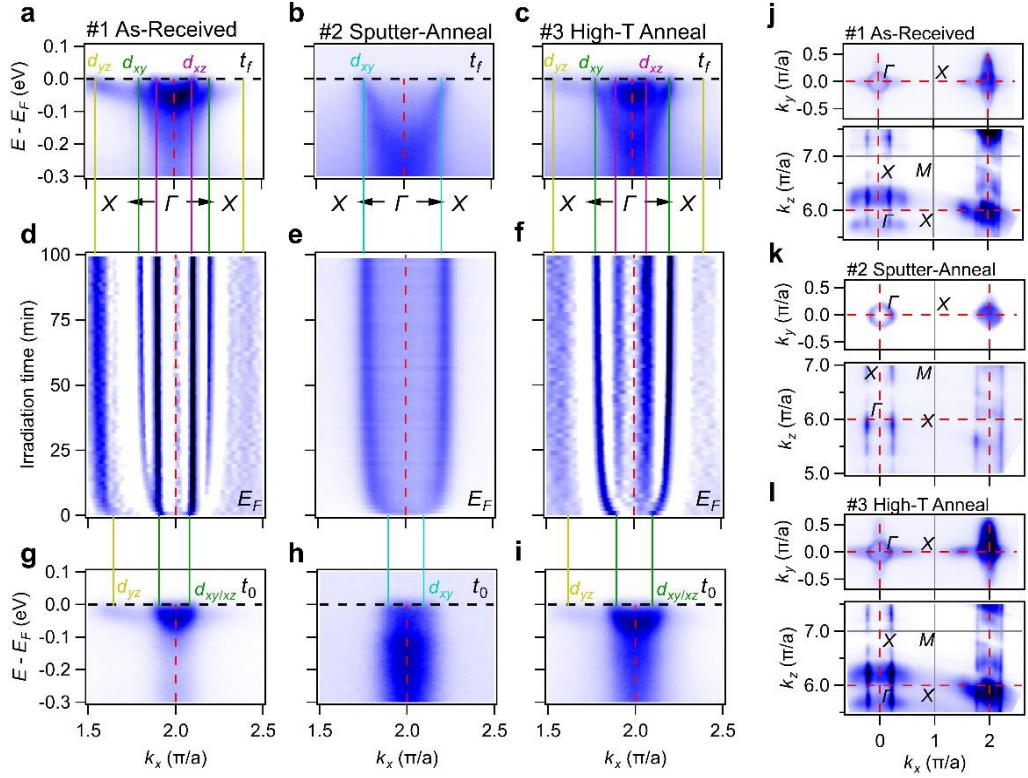

Figure 2. **a-c**, ARPES intensity cuts of as-received (stage #1), sputter-annealed (stage #2), and high temperature annealed (stage #3) STO wafers, respectively, after saturating the carrier density by irradiation ($t_f$). **d-f**, $k$-resolved ARPES maps at $E_F$ of stages #1 to #3, respectively, as a function of VUV irradiation time. **e** is an intensity map, and **d,f** are horizontal 2$^{nd}$ derivative maps. **g-i**, ARPES intensity cuts of stages #1 to #3 measured at a fresh spot and acquired within 2 minutes of VUV-irradiation ($t_0$), respectively. Green, pink, cyan and yellow lines mark the $k_F$ of $d_{xy}$, $d_{xz}$, $d_{xy}$ in #2, and $d_{yz}$ bands, respectively. **j-l**, Fermi surface maps of the $\Gamma XY$ plane (upper) by in-plane mapping and the $\Gamma XZ$ plane (lower) by $h\nu$-dependent mapping of stage #1 to #3, respectively. Grey solid lines mark the BZ boundary and red dashed lines show the high-symmetry lines. Figure **a-i** and the upper panel of **j-l** are measured at 85 eV. All figures are measured with circular polarized ($C+$) light.

**Reorganization of surface chemical composition and theoretical analyses**

Figures 3a and 3b present XPS spectra of the STO sample in the three stages by measuring the Sr $3d$ and Ti $3p$ and Ti $2p$ core levels with photon energies in the VUV (170eV; surface sensitive) and soft X-ray (750 eV; more bulk sensitive) ranges. In the regions of the spectra where core-levels of Ti are situated, the peaks at $E_B$~38 eV ($3p$) (Figure 3a) and at $E_B$~459 eV ($2p$) (Figure 3b) are due to Ti$^{4+}$ ions, while the peaks at

$E_B$~35 eV and at $E_B$~457 eV belong to the Ti$^{3+}$ ions. The overall shape of Ti$^{4+}$ is similar for all stages, while in stage #3 we have observed a minor amount of Ti$^{3+}$ as compared to the other two (Figure 3b). Yet a remarkable transformation happens with Sr 3$d$ core level of STO surface after sputtering and annealing (#2). While the data acquired by soft X-rays (Figure 3b) do not show the effect of the heat treatment, the VUV-XPS data reveals that the spectral weight of Sr at the surface is significantly raised (Figure 3a). Moreover, the same data show that the additional annealing (in #3) transforms the doublet peaks of the Sr 3$d$ core level into a multipeak (at least two doublets) structure, indicating the presence of chemically distinct Sr species after surface rearrangement.

By comparing the spectral weight of Sr 3$d$ and Ti 3$p$ core levels using VUV light (surface sensitive) for stages #1 and #2, it is evident that surface treatment yields an increased Sr content. Interestingly, with additional annealing (stage #3), the opposite trend was observed: the spectral weight of Sr 3$d$ decreases while the weight of Ti 3$p$ increases. To quantify this effect, we use the ratio of Sr and Ti ($I_{Sr}/I_{Ti}$) spectral weight (integral one) of the three stages obtained from the VUV and soft X-ray data, as shown in Figure 3c. We set the Sr/Ti ratio of as-received STO (stage #1) to unity and normalized the other values to it. As a result, the ratio $I_{Sr}/I_{Ti}$ of the surface region (VUV-extracted) increases by about 50% from stage #1 to stage #2 and decreases in stage #3. In contrast, the ratio $I_{Sr}/I_{Ti}$ of the bulk-like region (soft X-ray extracted) remains nearly unchanged. The obtained results suggest that the surface preparation alters the surface chemical composition from a nominally TiO$_2$-terminated surface (#1) to one with increased Sr content (#2). Additional annealing seems to restore the TiO$_2$ termination. S. N. Rebec et al. reported that the SrO layer deposited on TiO2-terminated STO yields a similar one-band electronic feature [27]. Furthermore, the same study showed that the $I_{Sr}/I_{Ti}$ ratio also increases around 50% for the SrO-capped sample [27], similar to our result presented in Figure 3c. All of these outcomes indicate that the surface chemical composition is crucial for modulating the electronic structures of STO surface.

To resolve the link between the STO surface composition and the band reconstruction, we employ theoretical calculations of TiO$_2$- and SrO- terminated STO to emulate the two possible final derivatives of surface preparation (Figure 3d,e). For the TiO$_2$-terminated STO slab, the calculation shows that the band splitting ($\Delta t_{2g}$) between $d_{xy}$ band bottom and $d_{yz}/d_{xz}$ band bottom is negligible (Figure 3f). This outcome is, indeed, in good agreement with band structures observed in stage #1 at $t_0$ (Figure

3h), indicating that the cubic symmetry is mostly preserved. In contrast, the calculations for SrO-terminated slabs show that the $d_{xy}$ band (derived from the first TiO$_2$ layer under SrO surface layer, which is highlighted in Figure 3g) shifts downward with ~150 meV, generating the splitting of $t_{2g}$ bands with tetragonal symmetry. Indeed, the ARPES data for stage #2 ($t_0$) reveals this type of 2D band, which is slightly shallower than the calculated one (Figure 3i). A possible explanation of this deviation is the larger effective mass of the observed band ($t_0$) caused by the strong EPI, which was not considered in the theoretical calculations. When the $E_F$ is set according to the experimental $k_F$ (Figure 3h, i), only the surface $d_{xy}$ band crosses $E_F$ for #2, while the other $t_{2g}$ bands remain unoccupied, hence showing that a single-band state arises in stage #2.

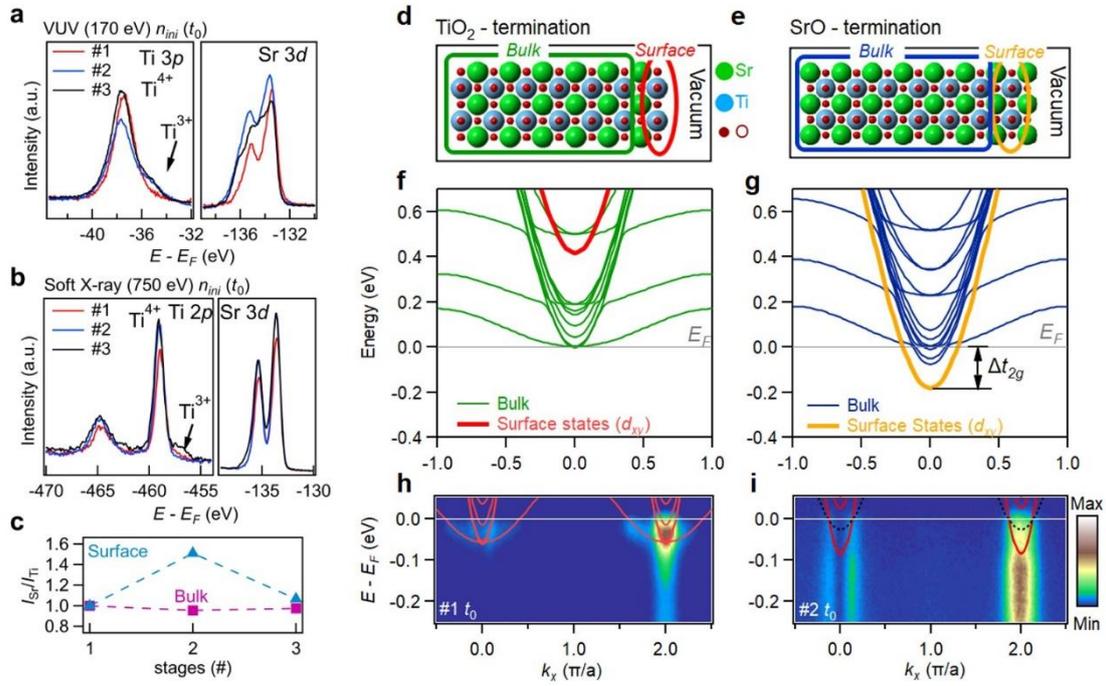

**Figure 3. a,b** Core-level of Ti 3$p$ and Sr 3$d$ orbitals measured at $h\nu$=170 eV at the initial carrier density of the three stages ($t_0$). **b**, Core-level of Ti 2$p$ and Sr 3$d$ orbitals measured at 750 eV at the initial carrier densities of the three stages ($t_0$). **c**, Calculated spectral weight ratio of Sr/Ti extracted from **a** and **b**, and normalized to the ratio of the as-received STO wafer (#1). **d-e**, Relaxed 2 × 2 × 7 STO slabs of TiO$_2$-termination and SrO-termination, respectively. Red/orange ellipses and dark green/blue squares mark the superficial TiO$_2$ layer and the bulk of two models, respectively. Green, blue, and red spheres represent the Sr, Ti, and O atoms, respectively. **f,g**, Calculated electronic band structures of TiO$_2$- and SrO- terminated STO models as presented in **a** and **b**,

respectively. The red and orange curve in **f** and **g** highlights the $d_{xy}$ band from the first TiO$_2$ layer. **h,i**, ARPES intensity cuts of as-received (#1) and sputter-annealed (#2) STO wafer at $t_0$, plotted with calculated band structures in **f** and **g**, respectively. The dashed black curves in **f** show the fitted $d_{xy}$ bands.

Intriguingly, a recent study of SrO-capped STO systems exhibits the absence of electronic states near $E_F$ [33]. This lack can arise due to the surface insensitive nature of soft X-ray incident light used for this study, which is unfavorable to detecting the $d_{xy}$ surface band in Sr-enriched STO. However, our ARPES and DFT data validate that the formed Sr-enriched STO (001) surface is characterized by the intrinsic splitting of the $t_{2g}$ states and the surface state, causing a pure 2D electronic structure. Therefore, the surface state with $d_{xy}$ character, which is first occupied when the system is doped, should be considered the main component of the 2DEG system [27,28,34].

**Wedge-potential in STO**

One of the key parameters of the STO band structures, shown in Figure 2, is the splitting of the $d_{xy}$ and $d_{xz}/d_{yz}$ bands ($\Delta t_{2g}$). E. B. Guedes et al., by combining DFT calculation and ARPES, established the link between atomic displacements of STO surface layers and $\Delta t_{2g}$ [23]. However, the trapped photo-generated electrons during APRES experiments can cause lattice distortion while free ones yield the electric field. Indeed, $\Delta t_{2g}$ alters during beam irradiation (Figure 2d,f) and directly correlates to carrier density.

Starting as a degenerate system (Figures 2g,i), the $d_{xy}$ and $d_{xz}/d_{yz}$ bands separate (Figure 2d, f) later during irradiation, resulting in 3D $d_{xz}/d_{yz}$ bands and 2D $d_{xy}$ subbands (Figure 2a and Figure 2c). The extracted Fermi momenta ($k_F$) and the energy splitting between $d_{xy}$ and $d_{xz}/d_{yz}$ bands ($\Delta t_{2g}$) as a function of irradiation time illustrate this process more clearly (see also Figure S3f). It has been extensively discussed in the literature that the STO-based 2DEG system experiences a wedge-like potential within the surface region [30, 35, 36]. This potential at the STO surface can be described using a quantum well model with the form, $V(z)=V_0+eFz$, where $F$ is the strength of the electric field in the direction perpendicular to the sample surface, and $e$ is the charge of the electron (Figure 4a). The quantized eigenenergies ($E_n$) of subbands at the surface region can be described by the equation 1[30, 36]:

$$E_n = V_0 + \left(\frac{\hbar}{2m_z^*}\right)^{1/3}\left[\left(\frac{3\pi}{2}\right)\left(n - \frac{1}{4}\right)eF\right]^{2/3} \quad \text{Eq. 1}$$

where $m^*_z$ is the effective mass along the field direction (perpendicular to the surface).

The band splitting of $t_{2g}$ bands, $\Delta t_{2g}$, can be calculated by the difference between the n=1 eigenenergies of $d_{xy}$ and $d_{xz}/d_{yz}$ bands as $\Delta t_{2g}=E_1^{xy}-E_1^{xz/yz}$ in the following way:

$$\Delta t_{2g} = 7.5 \times 10^{-7}\left[\left(\frac{m_e}{m_z^{*xz/yz}}\right)^{1/3} - \left(\frac{m_e}{m_z^{*xy}}\right)^{1/3}\right]F^{2/3} \quad \text{Eq. 2}$$

The relationship between $F$ and the carrier density ($n_{2D}$) can be describe in the following way [36]:

$$\frac{e}{2}n_{2D} = \int_0^F \varepsilon_0\, \varepsilon_r(F')dF' \quad \text{Eq. 3}$$

($\varepsilon_0$: the vacuum dielectric constant; $\varepsilon_r(F)$: is field-dependent dielectric constant of STO) From Eq. 3 it is possible to extract the electric field by using the carrier densities of each band taken from the ARPES data (See Figure S3 f,g, and Supplementary Sec. III). The effective mass $m^*_z$ of $d_{xy}$ and $d_{xz}/d_{yz}$ bands can also be estimated by ARPES.

Figure 4b shows the calculated and experimental values of $\Delta t_{2g}$ of stages #1 and #3 as a function of the carrier density, $n_{2D}$. The agreement between the observed and calculated values of $\Delta t_{2g}$ indicates that the electric field ($F$ or $\varepsilon_r(F)$) is caused by the accumulation of charges at the surface during the irradiation process.

All outcomes above and the results reported by E.B. Guedes [23], imply that the lattice distortion (stabilized by trapped electrons) cooperates vigorously with the field (generated by free electrons), conducting to the common properties of the 2DEG at the STO surface.

**EPI and effective mass**

The waterfall-like feature identified as an incoherent part of the band dispersion (see Figure 2g-h) is usually attributed to the polaronic electron-phonon interaction (EPI) in STO, which plays an essential role in modulating the physical properties of the 2DEG [37-39]. The EPI can be quantified by the quasi-particle (QP) residue, $Z_0 = I_{QP} / (I_{QP} + I_{hump})$, where the $I_{QP/hump}$ represents the integrated spectral weight of QP/hump of corresponding energy distribution curves (EDCs) [39,40].

Figure 4c displays the background-subtracted EDCs of stages #1-3 taken at the $k_F$ of the $d_{xy}$ bands for different carrier densities, $n(t)$, normalized by the QP peak's intensity. The peak-dip-hump line shape, which extends to higher binding energy (Figure 4c), relates to the multiple phonon modes interacting with electrons [39,40]. However, the EPI at $t_0$ (red curves in Figure 4c), related to intrinsic transport properties, exhibits a significant increase for stage #2 (a reduction of $Z_0$ from 0.55 in #1 to 0.15 in #2) and partially recovers after the sample has been additionally annealed ($Z_0$~0.35 in #3). Therefore, besides changes in the band topology (see Figure 2), surface engineering also modifies the EPI strength, affecting the carrier properties such as the effective mass, $m^*$ (from $0.6m_e$ in #1 to $1.8m_e$ in #2, and $0.8m_e$ in #3, See Figure S5). Figure 4d shows the EPI strength (through $Z_0$) as a function of the carrier densities of $d_{xy}$ bands. The positive (negative) $Z_0$ (EPI) behavior for different carrier densities is probably due to the screening suppressing the long-range Fröhlich polaron interaction [37,39]. It is important to note that the spectral weight of the peak-dip-hump structures undergoes a continuous decrease during irradiation, indicating a weakening of the EPI for all stages (#1 to #3).

In weak coupling and long-range electron-phonon interactions, the relationship between effective mass and quasiparticle residue ($Z_0$) is defined by the Fröhlich model, where the effective mass of the electrons is enhanced due to EPI: $m^*/m_0=1/(1-\alpha/6)$, $m_0$ is the bare band mass, and $\alpha$ is the coupling strength [41,42], which can be estimated from $Z_0$ by a diagrammatic quantum Monte Carlo study (presented in Figure 4d with gradient background and the right axis) [43]. In Figure 4e, we plot both effective mass fitted from band dispersion (fitting details are presented in Supplementary Section III and IV) and calculated one from $Z_0$ by weak-coupling Fröhlich model of stages #1 as a function of carrier density. Our data show good agreement between the fitted $m^*$ and calculated values for stages #1 and #3 for all observed carrier densities. However, for stage #2 this concurrence is valid only for the high doping regime, which is consistent with the weak coupling regime (See Figure 4d, and more details are shown in Figure S6). For the low-doping regime of stage #2, with $Z_0 < 0.3$ and $\alpha > 3$ (Figure 4d), a strong coupling model we used for the approximation established by R. P. Feynman [44], $m^*/m_0=0.0232\alpha^4$, which reproduces the effective mass more appropriately, as shown in Figure 4e.

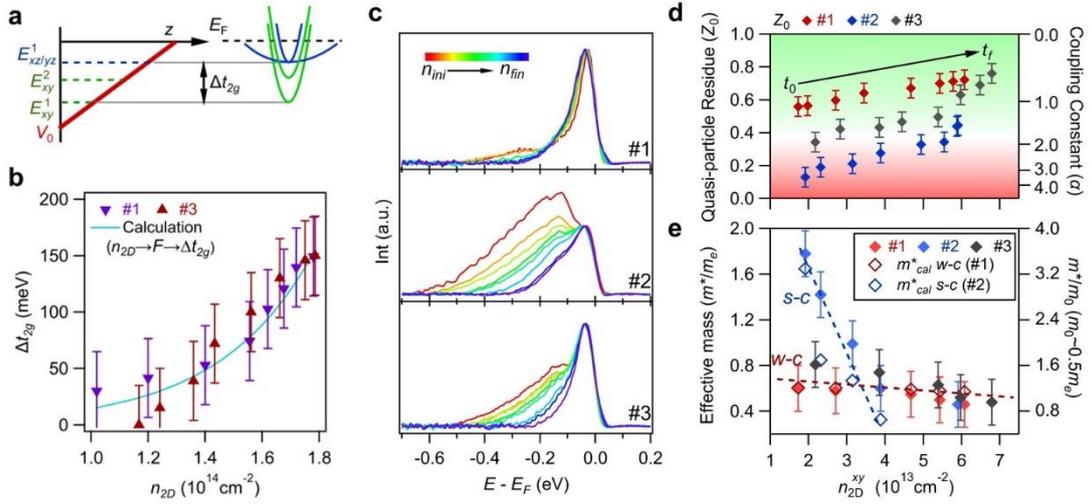

Figure 4. **a**, Wedge potential and band structure at the surface of STO. **b**, Calculated energy differences between $d_{xy}$ and $d_{xz}/d_{yz}$ bands ($\Delta t_{2g}$) as a function of carrier density, and experimental $\Delta t_{2g}$ extracted from ARPES data of stage #1 and #3. **c**, Energy distribution curves of stage #1 to #3 at different carrier densities at $k=k_F^{xy}$. **d**, Calculated quasiparticle residue ($Z_0$) from the EDCs (presented in panel **a**) as a function of carrier density of $d_{xy}$ bands, $n_{2D}^{xy}$. The gradient background and right axis indicate the transition between weak (light green) to strong (light red) coupling strength (reproduced from ref. 42). **e**, Fitted effective masses of $d_{xy}$ bands of all three stages as a function of $n_{2D}^{xy}$. The empty marks represent the calculated effective mass from coupling strength ($\alpha$) by weak-coupling, $m^*/m_0=1/(1-\alpha/6)$, and strong coupling, $m^*/m_0=0.0232\alpha^4$, of Fröhlich polarons, respectively.

Our data and analyses in Figures 4e and 4d show that the EPI transit from the weak-coupling (stage #1) to the strong-coupling regime (stage #2). Further, the verified inverse relation between the EPI strength and the effective mass of all three stages shows the softening of polarons caused by increasing carrier densities [37].

## Summary and Outlook

Employing systematic XPS measurements, we show that combining Ar sputtering and UHV annealing modifies the STO chemical composition, transforming nominally $TiO_2$- terminated to SrO-enriched surface. Utilizing ARPES and DFT calculations validates that the formed Sr-enriched STO (001) surface is characterized by the intrinsic splitting of the $t_{2g}$ states while the $d_{xy}$ surface state yields a pure 2D electronic structure.

The observed single-band electronic phase is air-stable (Supplementary, Sec.II), showing the potential to service in designing novel devices.

The additional UHV annealing at moderate temperature eradicates Sr from the surface and partially recovers the nominally $TiO_2$-terminated character, leading to the Lifshitz-like transition in STO (from one band to three bands metallicity).

Our work describes a straightforward method for varying the surface chemical composition, which, combined with VUV-irradiation, efficiently modulates the electronic structures of the $t_{2g}$ band, doping, and electron-phonon interaction in STO.

# METHODS

## Sample Preparation

In this study, as-received commercially available 0.5% Nb doped nominally $TiO_2$-terminated $SrTiO_3$ wafers (SurfaceNet GmbH) are used with a miscut within 0.2° to the nominal (001) surface.

During ARPES measurements, multiple surface engineering procedures are applied to the as-received STO wafer, including sputtering and annealing. The sputtering process is conducted under an argon atmosphere with a pressure of $2\times10^{-6}$ mbar. The voltage is set as 1kV and the STO wafers are 45° facing the ion beam. The vacuum annealing process is conducted under an ultra-high vacuum better than $2\times10^{-8}$ mbar. The annealing temperatures are read through an infrared thermometer. The detailed sample treatment routes are shown in Supplementary, Sec. I. In our studies, we reproduce similar results of band modulation in at least another three samples, which are shown in Supplementary Sec. V.

## Angle-resolved photoemission spectroscopy

All the ARPES and XPS data presented were measured at the ULTRA endstation at the Surface/Interface Spectroscopy (SIS) beamline of the Swiss Light Source. The data were acquired with a Scienta Omicron DA30L hemispherical analyzer. The energy and angular resolution are better than 20 meV and 0.1°. The measurements were performed at a temperature of 20 K in a base pressure better than $1\times10^{-10}$ Torr. The un-irradiated results are measured by moving the samples to un-irradiated areas.

## DFT calculation

The density functional theory calculations were performed within the Perdew–Burke–Ernzehof generalized gradient approximation (PBE-GGA) [45], using projector augmented wave (PAW) potentials [46], as implemented in the Vienna *ab initio* Simulation Package (VASP) [47,48]. In addition, the DFT+U functional of Liechtenstein et al. [49] was employed with U = 5 eV and J = 0.64 eV, as similarly performed in ref. [50]. A basis set of 500 eV were used, and the structures were relaxed until the forces on atoms were less than 0.01 eVÅ$^{-1}$. The relaxation of the atomic positions was done using a 4 × 4 × 1 k-mesh, whereas the band structures were evaluated using an 8 × 8 × 2 k-points set.

**Acknowledgements**

This work was supported by the Swiss National Science Foundation (SNF), No. 200021_182695 and No. 200021_188413. W.H.B acknowledges the support from the Brazilian agencies CNPq, FAPEMIG, and CAPES, as well as the CENAPAD-SP, CESUP (UFRGS), and the National Laboratory for Scientific Computing (LNCC/MCTI, Brazil) for providing HPC resources of the SDumont supercomputer, which have contributed to the research results, URL: http://sdumont.lncc.br.

**Author contributions**

M.R. designed the concept and experiments with N.P. and H.L. H.L., E.B.G. A.C. and M.R. performed the ARPES experiment with the help from R.T.D., N.C.P. and M.S. W.H.B. performed the DFT calculations. H. L. processed the ARPES data. H.L. and M.R. wrote the manuscript assisted by E.B.G and A.C and with the help and useful discussions with all other authors.

**Competing interests:** The authors declare that they have no competing interests.

**Data and materials availability**: All data needed to evaluate the conclusions in the paper are present in the paper and/or the Supplementary Materials. Materials and additional data related to this paper may be requested from the authors.